\documentclass{article}
\usepackage{spconf,amsmath,graphicx}

\usepackage{amsmath} % some math-related packages, not sure which of them are necessary
\usepackage{amssymb}
\usepackage{amsfonts}
\usepackage{subfigure}
\usepackage{lipsum, soul,xcolor} 
\usepackage{url}
\title{ROBUST SPEAKER EXTRACTION NETWORK BASED ON \\ ITERATIVE REFINED ADAPTATION}
\name{Chengyun Deng, Shiqian Ma, Yi Zhang, Yongtao Sha, Hui Zhang, Hui Song, Xiangang Li}

\address{Didi Chuxing, Beijing, China}
\begin{document}
%\ninept
%
\maketitle
\begin{abstract}
Speaker extraction aims to extract target speech signal from a multi-talker environment with interference speakers and surrounding noise, given the target speaker’s reference information. Most speaker extraction systems achieve satisfactory performance on the premise that the test speakers have been encountered during training time. Such systems suffer from performance degradation given unseen target speakers and/or mismatched reference voiceprint information. In this paper we propose a novel strategy named Iterative Refined Adaptation (IRA) to improve the robustness and generalization capability of speaker extraction systems in the aforementioned scenarios. Given an initial speaker embedding encoded by an auxiliary network, the extraction network can obtain a latent representation of the target speaker, which is fed back to the auxiliary network to get a refined embedding to provide more accurate guidance for the extraction network. Experiments on WSJ0-2mix-extr and WHAM! dataset confirm the superior performance of the proposed method over the network without IRA in terms of SI-SDR and PESQ improvement.
\end{abstract}
%open-set condition 
\section{Introduction}
\label{sec:intro}

% Recent studies on time-domain audio separation network (TasNet) with encoder-separator-decoder architecture have seen significant progress in speaker-independent speech separation tasks, such as Conv-TasNet \cite{luo2019conv}, DPRNN-TasNet \cite{luo2020dual} and DPTNet \cite{chen2020dual}. Compared with time-frequency approaches \cite{li2018cbldnn, wang2018alternative}, time-domain separation methods can bypass the difficulty of phase reconstruction and avoid the long latency effect caused by the calculation of short-time Fourier transform (STFT) \cite{luo2019conv}. 

Auditory attention allows human to focus on a specific speaker in a crowd, which is also called the cocktail party effect \cite{getzmann2017switching}. Speaker extraction aims to extract the target speaker signal in a challenging acoustic environment according to the reference voiceprint information. Speaker extraction plays an important role in improving the intelligibility of speech. 

Speaker extraction algorithms can be roughly divided into three categories, i.e., time-frequency (T-F) approaches, complete time-domain approaches and hybrid approaches. T-F methods aim to encode the target speaker information using T-F features. They estimate the target speaker's magnitude spectrum guided by the speaker embedding \cite{wang2018voicefilter, delcroix2018single, xiao2019single}, and then rebuild the waveform via inverse short-time Fourier transform by combining the extracted magnitude spectrum with potentially mismatched phase from the mixture, such as Voicefilter \cite{wang2018voicefilter} and  SBF-MTSAL-Concat \cite{xu2019optimization}. Hybrid methods are then proposed integrating a T-F based speaker embedding with a time-domain speech separation network, such as SpEx \cite{xu2019time}. Furthermore, to avoid the mismatch of latent feature space between speech encoder and speaker encoder, a complete time-domain method SpEx+ \cite{ge2020spex+} is proposed. The mixture and reference speech are represented by CNN based multi-scale weight-shared speech encoders.

Numerous speaker extraction systems achieve good performance with seen target speakers and matched reference voiceprint guidance (we label it as closed condition). However, realistic circumstances with a large number of unseen target speakers and/or mismatched reference voiceprint information (we label it as open condition) have an adverse effect on most speaker extraction systems. How to improve the robustness and generalization capability of speaker extraction networks in open conditions is still an open question yet. 

Motivated by this phenomenon, we propose an Iterative Refined Adaptation strategy to make the extraction procedure more reliable. Inspired by SpEx+ \cite{ge2020spex+}, we present a complete time-domain speaker extraction algorithm that integrates DPRNN-TasNet \cite{luo2020dual} and time-domain residual network (TD-ResNet), labeled as DPRNN-Spe. For the purpose of refining speaker embedding in case of mismatched reference, we furtherly apply IRA, labeled as DPRNN-Spe-IRA. Firstly, We use the original speaker embedding encoded by TD-ResNet to extract target latent representations from the mixture. Secondly, TD-ResNet re-encodes the latent representations to obtain a refined text-dependent speaker embedding, which is then combined with the original one with weights to form a new embedding through a linear layer. We extract the target speech using the new embedding correspondingly. Finally, we reconstruct the waveform of the target speaker by the decoder. Experiments on WSJ0-2mix-extr \cite{xu2018single} show that DPRNN-Spe-IRA achieves 1.05 dB SI-SDR improvement over DPRNN-Spe baseline, and 0.58 dB SI-SDR improvement over the state-of-the-art SpEx+, using single-scale speech encoder with filter length of 16 samples. Moreover, DPRNN-Spe-IRA yields better extraction performance in noisy scenarios as well.

We will first give an overview of DPRNN-Spe framework in Section 2 and then go into details about IRA in Section 3. In Section 4 we present our experimental results and then come to conclusions in Section 5.

\vspace{-2mm}
\section{DPRNN-Spe Architecture}
\label{sec:DPRNN-Spe-architecture}

Single-channel speaker extraction system can be formulated in terms of extracting speech of target speaker from the mixtures $x(t)$ with $C$ sources (one target speaker $s_{target}(t)$ and $C-1$ interference sources $s_{i}(t)$).

\begin{figure}[t!]
\centering
\includegraphics[scale = 0.22]{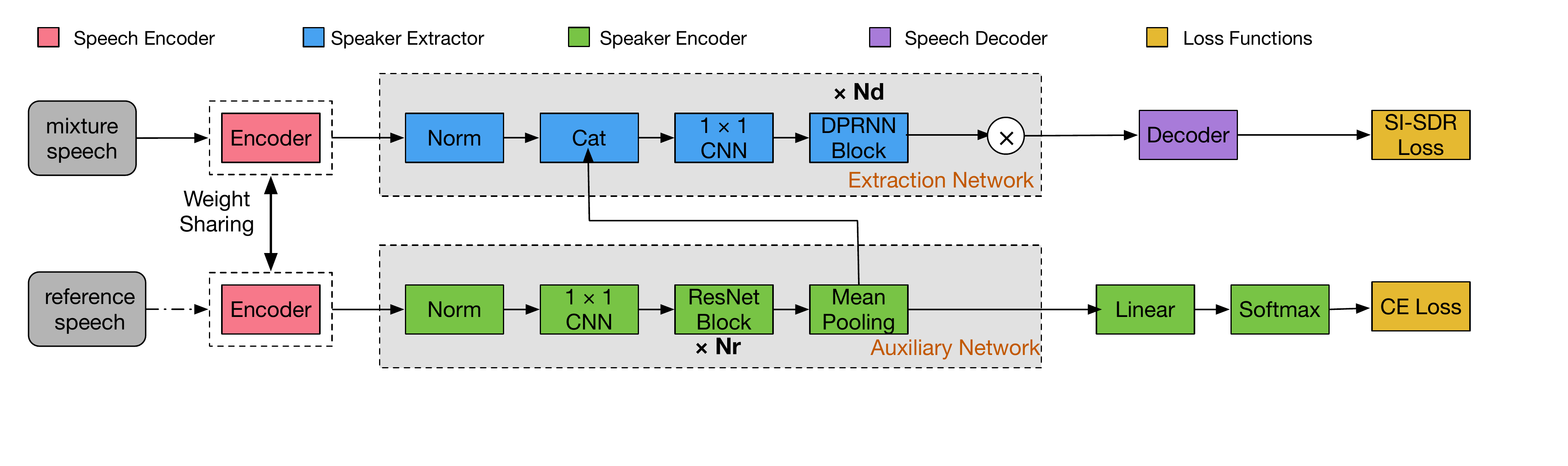}
\vspace{-10mm}
\caption{The block diagram of the speech extraction system labeled as DPRNN-Spe. }
\label{fig:DPRNN-Spe}
\vspace{-3mm}
\end{figure}
\noindent 

% \begin{equation}
% \label{mixture}
% \begin{aligned} 
% x(t)=s_{target}(t) + \sum_{i=1}^{C-1}s_{i}(t)
% \end{aligned}
% \end{equation}

According to \cite{luo2020dual}, dual-path recurrent neural network (DPRNN) can make use of global information and achieve superior performance with smaller model size comparing with temporal convolutional network (TCN). Thus we use DPRNN to build our extraction system. As shown in Figure \ref{fig:DPRNN-Spe}, the proposed DPRNN-Spe system consists of twin speech encoders, a DPRNN based extraction network, a ResNet based auxiliary network and a decoder. The difference between DPRNN-Spe and SpEx+ lies in three aspects, (i) we use single-scale rather than multi-scale encoders, (ii) the speaker embedding is concatenated just once instead of repeatedly before each single DPRNN block, and (iii) we use DPRNN blocks rather than TCN blocks for the extraction network. 

\vspace{-2mm}
\subsection{Twin speech encoders}
\label{encoder}
% \vspace{-2mm}
The mixture speech $x(t)$ and reference speech $r(t)$ can be transformed to latent representations $Mix\_E$ and $Spe\_E$ by weight-shared 1-D CNNs \cite{ge2020spex+} with length $L$ and stride $L/2$, respectively.

\begin{figure*}[h!]
  \centering
  \includegraphics[scale = 0.30]{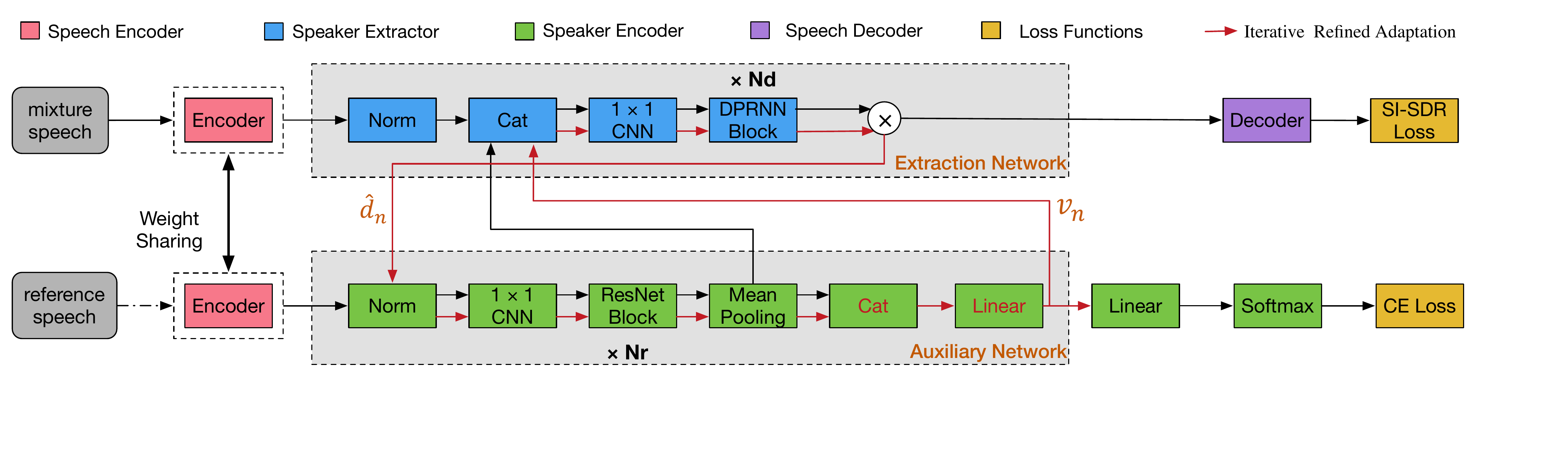}
  \vspace{-10mm}
  \caption{The block diagram of the speech extraction system with iterative refined adaptation labeled as DPRNN-Spe-IRA. The speaker embedding $v_n$ encoded by the auxiliary network and the representation of target speaker $\hat{d}_n$ extracted by the extraction network iteratively refine and adapt to each other.}
  \label{fig:IRA}
  \vspace{-3mm}
\end{figure*}
\noindent

% \begin{equation}
% \label{embedding}
% \begin{aligned} 
% v = \mathbf{A}(Spe\_E)
% \end{aligned}
% \end{equation}

\vspace{-2mm}
\subsection{Auxiliary network}
\label{aux}
We extract a highly discriminative embedding $v = \mathbf{A}(Spe\_E)$ from $Spe\_E$,  where $\mathbf{A}(\cdot)$ represents the auxiliary network. Similar to SpEx+, we use ResNets as the core of the auxiliary network.
% ResNets are commonly used in speaker recognition tasks \cite{xie2019utterance, li2018deep}.
% Our ResNet block has similar architecture but different parameter configuration from SpEx+. Our auxiliary network is with smaller size since we don't use multi-scale encoders. 
% Our speaker embedding dimension is set to 128 rather than 256. 

% As the filter length of the encoder increases, the total information represented by each frame of $Spe\_E$ is less, although the information is more refined. The modeling of speaker embedding requires information of sufficient frames, not too detailed information. We need to change the stride and kernel size of each ResNet's Maxpooling layer to adapt to the modeling accuracy of the speaker embedding. 

Following the ResNet blocks and the mean-pooling operation, a linear fully-connected (FC) layer and a softmax function are used in the speaker classification task. In other words, the speaker classification is a sub-task in the multi-task learning of DPRNN-Spe. A cross-entropy loss $\pounds _{CE}$ \cite{ge2020spex+} is used for speaker classification.
% \begin{equation}
% \label{mixture}
% \begin{aligned} 
% \pounds _{CE}=-\sum_{i=1}^{N_{s}}I_{i}log(\hat{p_{i}})
% \end{aligned}
% \end{equation}
% where $N_{s}$ is the total number of speakers in training data, $I_{i}$ is the true class label for the $i^{th}$ speaker, and $\hat{p_{i}}$ is the corresponding predicted probability.

\vspace{-2mm}
\subsection{Extraction network}
The speaker embedding is repeatedly concatenated along the feature dimension of the normalized $Mix\_E$ before the first DPRNN block. Similar to DPRNN-TasNet \cite{luo2020dual}, our extraction network $\mathbf{E}$ includes 6 DPRNN blocks, where BLSTM \cite{hochreiter1997long} is used as the intra- and inter-chunk RNNs, and each direction has 128 hidden units. The speaker extraction network predicts a latent representation $\hat{d}$ of the target speaker by learning a mask $m$ in this latent space,
\begin{equation}
\label{extraction}
\begin{aligned} 
m = \mathbf{E} ([v:norm(Mix\_{E})]) \\
\hat{d} = Mix\_{E} \odot m
\end{aligned}
\end{equation}
where $[:]$ is the concatenate operation and $\odot$ denotes element-wise multiplication.
% The representation of the target speaker can be calculated by applying the mask to $ Mix\_{E}$,

\vspace{-2mm}
\subsection{Speech decoder}
\label{decoder}
Finally, the overlapping-segment waveform of the target source $\hat{\mathbf{s}}_t $ is reconstructed by the decoder consisted of 1-D transposed convolution. The overlapping reconstructed matrix generates the final waveform $\hat{s}_{target}(t)$ by overlap and sum operation.
% \begin{equation}
% \label{mixture}
% \begin{aligned} 
% \hat{\mathbf{s}} = \hat{d}V
% \end{aligned}
% \end{equation}
% where $V\in \mathbb{R}^{N \times L}$ is the decoder basis functions matrix. The overlapping reconstructed matrix generates the final waveform $\hat{s}_{target} \in \mathbb{R}^{1\times T} $ by overlap and sum operation.

\vspace{-2mm}
\subsection{Multi-Task Learning}
\label{losses}

The main objective of DPRNN-Spe training is to maximize the scale-invariant source-to-distortion ratio (SI-SDR) \cite{le2019sdr} of the target speaker, with the assistance of high-quality speaker embedding. 
% SI-SDR is defined as:
% \begin{equation}
% \label{mixture}
% \begin{aligned} 
% \left\{\begin{matrix}
% s_{target}:= \frac{<\hat{s},s>s}{\left \| s \right \|^{2}}\\ 
% e_{noise}:=\hat{s}-s_{target}\\ 
% SI\text{-}SDR:=10log_{10}\frac{\left \| s_{target} \right \|^{2}}{\left \| e_{noise} \right \|^{2}}
% \end{matrix}\right.
% \end{aligned}
% \end{equation}
% where $\hat{s}$ and $s$ are the predicted and original clean sources, respectively, and $\left \| s \right \|^{2} = <s, s>$ denotes the signal power. 

Meanwhile, $\pounds _{CE}$  is used as a regularization to guide the auxiliary network training. Hence the multi-task learning loss of DPRNN-Spe is defined as:
\begin{equation}
\label{SI-SNR}
\begin{aligned} 
\pounds _{DPRNN\_Spe}  = \pounds _{SI\text{-}SDR} + \lambda \pounds _{CE}
\end{aligned}
\end{equation}
where $\lambda$ is a hyper-parameter used to balance $\pounds _{SI\text{-}SDR}=-SI\text{-}SDR$ and $\pounds _{CE}$.

\section{Iterative Refined Adaptation}
\label{IRA_main}

\vspace{-2mm}
\subsection{Robustness problems}
\label{problems}
Speaker extraction systems often yield inferior results in open conditions. Apart from unseen target speakers, mismatch problems may also exist. Speaker embedding of one person may change due to factors such as age, physical health, mood, speaking rate, etc. Besides, the mixture and reference speech may be recorded in different acoustic environments or different channels, making the reference voiceprint information misleading to some extent. That is to say, speaker embedding would vary over time and environments. Even if it is tolerable in speaker recognition, it may cast negative effects on speaker extraction tasks. Furthermore, text-independent speaker embedding may lead to information redundancy compared to text-dependent embedding.

\vspace{-2mm}
\subsection{Formulation of iterative refined adaptation}
\label{formulation}

We propose a training strategy called IRA to eliminate such adverse effects. Suppose we have an initial mismatched reference $r$ and a main extraction function $\boldsymbol{F}$ as well as an auxiliary function $\boldsymbol{A}$ of any kind with input mixture $y$. We obtain a rough extraction result $\hat{x}_0$ and a prior condition for $\boldsymbol{F}$ on the basis of original condition $a_0 = \boldsymbol{A}(r)$. Then we feed $\hat{x}_0$ back to the auxiliary function to produce a refined reference information $a_1$, which is then fed back into $\boldsymbol{F}$ to produce a more accurate result $\hat{x}_1$. After $n$ $(n=1,2,...)$ times of feedback and modification, the network can deliver a more matched condition $a_{n}$ and a more accurate result $\hat{x}_n$:

\begin{equation}
\label{IRA_formulation}
\begin{aligned}
a_{n} = a_{n-1} + \mu \boldsymbol{A} (\hat{x}_{n-1}) \\
\hat{x}_n = \boldsymbol{F} (y \mid a_{n})
\end{aligned}
\end{equation}
where $\mu$ is a scaling parameter.

\vspace{-2mm}
\subsection{Speech extraction system with IRA}

It is noteworthy that IRA can be applied to any extraction systems easily. In this paper, we validate its effectiveness using our speaker extraction network DPRNN-Spe. We extend DPRNN-Spe to DPRNN-Spe-IRA illustrated in Figure \ref{fig:IRA}. The difference lies in the auxiliary network, which has an additional concatenate operation and an additional FC layer following the mean pooling operation. 

First of all, we use the original reference speech to obtain a initial embedding $v$ ($v_0$). Then, we get the estimated representation $\hat{d}$ ($\hat{d}_0$) of the target speaker from the extraction network, with input features the embedding $v_0$ and $Mix\_E$. Next, we feed $\hat{d}_0$ back to the auxiliary network to produce a new embedding $\mathbf{A}(\hat{d}_0)$, and concatenate it together with $v_0$. The additional FC layer is used to transform the feature dimension of the above embedding back to the that of $v_0$. And then this refined embedding $v_1$ is fed back into the extraction network to produce a better extracted representation $\hat{d}_1$. That is, with the information of the auxiliary network and the extraction network adapting mutually and repeatedly $n$ $(n=1,2,...)$ times, we can get a more matched embedding $v_n$ and a more accurate mask $m_n$ (leads to a better representation $\hat{d}_n$), as,
\begin{equation}
\label{FC}
\begin{aligned}
v_{n} = ([v_{n-1}: \mathbf{A}(\hat{d}_{n-1})])W + B \\
m_n = \mathbf{E} ([v_n:Mix\_{E}]) \\
\hat{d}_n = Mix\_{E} \odot m_n 
\end{aligned}
\end{equation}
where $W$ and $B$ are the weights and bias of the FC layer, respectively.

In particular, we replace batch normalization \cite{ioffe2015batch} with global layer normalization \cite{ba2016layer} in ResNet blocks for different data distribution of $Spe\_E$ and  $\hat{d}_n$.

\section{Experimental Evaluation}
\label{ssec:setup}
% \vspace{-1mm}

\vspace{-2mm}
\subsection{Dataset}
\label{data}
%  WSJ0-2mix-extr contains 30 hours of training data with 101 persons, 10 hours of validation data with the same 101 persons and 5 hours of evaluation data with 18 different persons. 

We evaluate the speech extraction performance using WSJ0-2mix-extr dataset \cite{xu2018single} and WHAM! dataset \cite{wichern2019wham}. WSJ0-2mix-extr is noise-free, and WHAM! contains numerous real ambient noise samples. Sample rate is 8kHz. In WSJ0-2mix-extr, the first speaker $s1$ is the target speaker and reference speech of the target speaker is randomly selected in clean speech except for the one in the mixture of same person. We simulate a noisy speaker extraction dataset based on WHAM!, as well. We choose both speakers as target speakers, and the selection of reference speech is the same as the WSJ0-2mix-extr.

% The mixture speech are sum of a pair of utterances form different speakers at random signal-to-noise ratios (SNR) that uniformly between 0 dB and 5 dB.

% WSJ0-2mix-extr is max version, but WHAM! is min version in this paper.
% Because of additive noise, average SI-SDR of test is -4.50 dB. 
% Filenames of all reference speech can be found at GitHub. 

\vspace{-2mm}
\subsection{Training and evaluation setup}
\label{setup}
For DPRNN-Spe, we use the same encoder and decoder design as in \cite{luo2020dual}. The number of ResNet blocks in auxiliary network is set to 3, and dimension of the speaker embedding is set to 128. For DPRNN-Spe-IRA, the additional FC layer has 128 linear node. The hyper-parameter that control the CE loss of the speaker classifier is set as $\lambda=0.5$. 

All models are trained using Adam optimizer \cite{kingma2014adam} for 100 epochs on 4-second long segments with an initial learning rate of 0.0005 and using a batchsize of 12 for $L=16$ and 8 for $L=8$. The learning rate is divided by 2 if the validation loss does not improve for 2 epochs. 

\begin{table}[t!]
\caption{SDRi (dB), SI-SDRi (dB) and PESQ of extracted speech for different systems on the WSJ0-2Mix-Extr. Best scores are highlighted in \textbf{bold}. $L$ is the filter length of the encoder.}
% \vspace{-2mm}
\label{results}
\centering 
\small
\scalebox{1.00}{
\begin{tabular}{lcccc}
\hline
\multicolumn{1}{l|}{\textbf{Algorithms}} & \textbf{Params} & \textbf{SI-SDRi} & \textbf{SDRi} & \textbf{PESQ}  \\ \hline
% \multicolumn{1}{l|}{SpeakerBeam \cite{delcroix2018single}} &-  &5.720 & - &- & - \\ \hline
% \multicolumn{1}{l|}{SBF-MTSAL-Concat \cite{xu2019optimization}} &8.900  &8.100 & - &- & - \\ \hline
\multicolumn{1}{l|}{Mixture} & - & 0.00 & 0.00 &2.31 \\ \hline
\multicolumn{5}{c}{$L=(20,80,160)$}   \\ \hline
\multicolumn{1}{l|}{SpEx \cite{xu2019time}} & 10.80M  & 14.18 & 14.55 &3.36 \\ \hline
\multicolumn{1}{l|}{SpEx+ \cite{ge2020spex+}} &13.30M  &15.70  &15.94 & 3.49 \\ \hline
\multicolumn{5}{c}{$L=16$}   \\ \hline
\multicolumn{1}{l|}{DPRNN-Spe} &2.91M  & 15.23 &15.48 &3.42  \\ \hline
\multicolumn{1}{l|}{DPRNN-Spe-IRA} &2.94M &\textbf{16.28} &\textbf{16.53} &\textbf{3.53} \\ \hline
\multicolumn{1}{l|}{DPRNN-Spe-2IRA} &2.94M &\textbf{16.45
} &\textbf{16.70} &\textbf{3.54} \\ \hline
\multicolumn{5}{c}{$L=8$}   \\ \hline
\multicolumn{1}{l|}{DPRNN-Spe} &2.90M  & 15.94 &16.27 &3.50 \\ \hline
\multicolumn{1}{l|}{DPRNN-Spe-IRA} &2.94M  &\textbf{17.50} &\textbf{17.73} &\textbf{3.62} \\ \hline
% \multicolumn{5}{c}{$L=4$}   \\ \hline
% \multicolumn{1}{l|}{DPRNN-Spe} &2.90M  & - &- &-  \\ \hline
% \multicolumn{1}{l|}{DPRNN-Spe-IRA} &2.94M  &\textbf{-} &\textbf{-} &\textbf{-} \\ \hline
% \multicolumn{1}{l|}{DPRNN-Spe} &2.904  & - &- &- &- \\ \hline
% \multicolumn{1}{l|}{DPRNN-Spe-IRA} &2.937  &\textbf{-} &\textbf{-} &\textbf{-} &\textbf{-}\\ \hline
\end{tabular}
}
\vspace{-3mm}
\end{table}

% \begin{figure}[t!]
% \centering
% \includegraphics[scale = 0.22]{SI-SDR_SD.png}
% \vspace{-2mm}
% \caption{Mean SI-SDR(dB) of validation data (closed condition) and test data(open condition) on the  WSJ0-2mix-extr. Error bars indicate standard error.}
% \label{fig:SD}
% \vspace{-3mm}
% \end{figure}
% \noindent

\begin{figure}[t!]
\centering
\subfigure[validation data]{
\includegraphics[scale = 0.20]{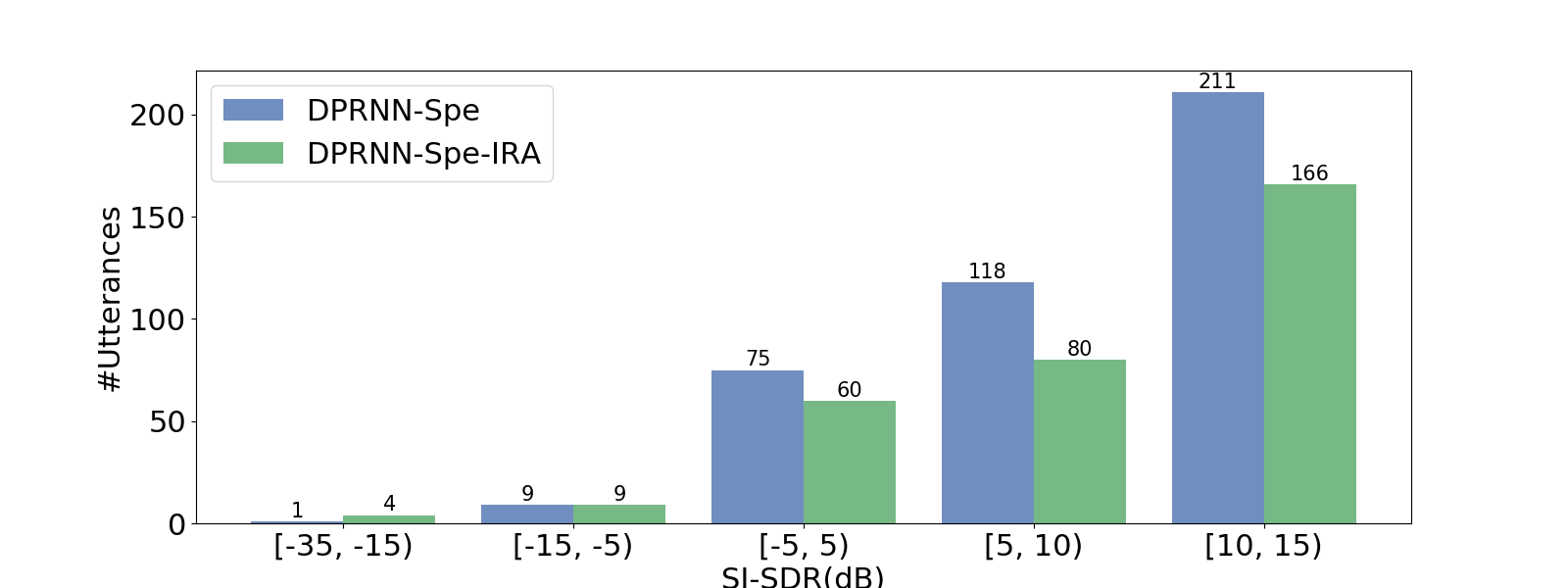}}
% \vspace{-1.5mm}
\subfigure[test data]{
\includegraphics[scale = 0.20]{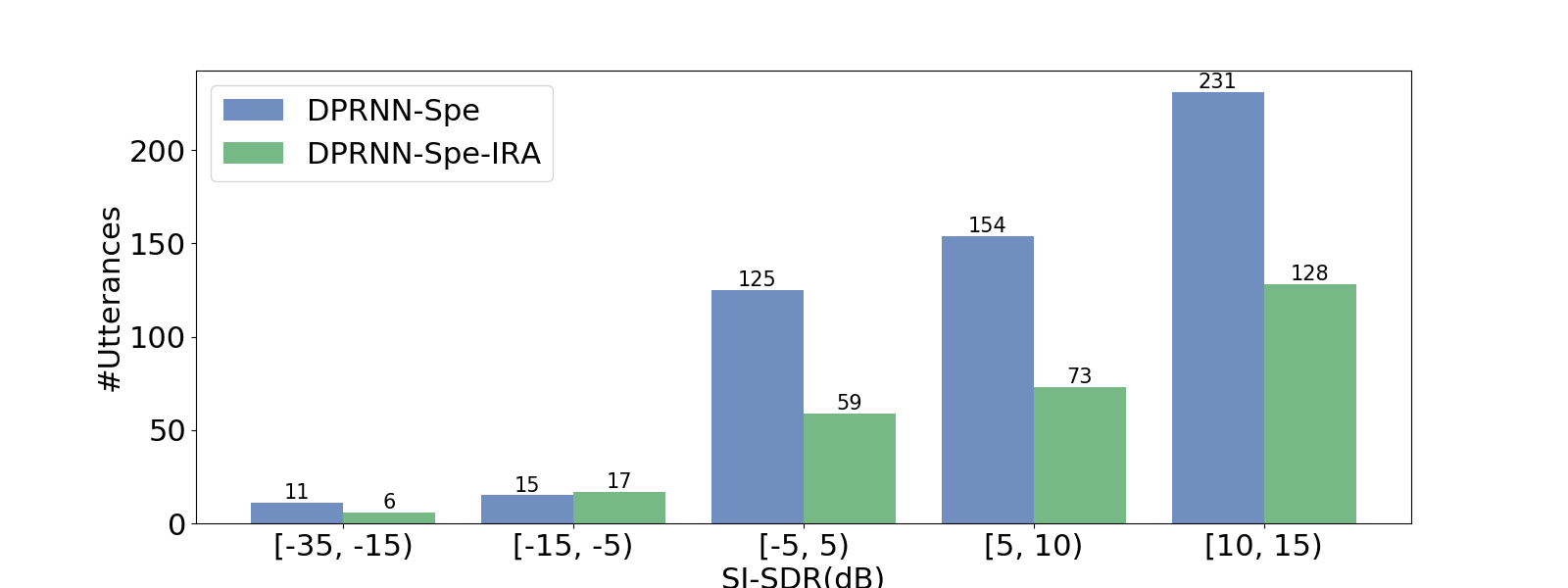}}
% \vspace{-1.5mm}
\caption{Distributions of the number of utterances with SI-SDR lower than 15dB. The smaller number of utterances with low SI-SDR suggests better performance of extraction.}
\label{fig:Distribution}
\vspace{-3mm}
\end{figure}
\noindent

We use the SI-SDR improvement (SI-SDRi), signal-to-distortion ratio improvement (SDRi) \cite{vincent2006performance} and perceptual evaluation of speech quality (PESQ) \cite{rix2001perceptual} as objective measures of extraction accuracy.

%\vspace{-3mm}
% \subsection{Results}
% \label{sec:results}

\vspace{-2mm}
\subsection{Comparative study on WSJ0-2mix-extr}
\label{wsj0-2mix-extr}

The results for the different algorithms are shown in Table \ref{results}. SpEx+ is the state-of-the-art method of speaker extraction. In $L=16$ case, We notice that the proposed DPRNN-Spe algorithm has matching performance with SpEx+ without multi-scale speech encoders and a smaller auxiliary network. DPRNN-Spe-IRA yields an absolute SI-SDR improvement over DPRNN-Spe of up to 1.05 dB, an absolute PESQ improvement of 0.11. DPRNN-Spe-IRA significantly outperforms previous state-of-the-art SpEx and SpEx+ with relative improvements of 14.81\% and 3.69\% in terms of SI-SDR, respectively. DPRNN-Spe-IRA results in a 6.91\% relative improvement in terms of SI-SDR comparing with DPRNN-Spe, and DPRNN-Spe-2IRA (that is, DPRNN-Spe uses IRA twice) results in a 8.03\% relative improvement. That is, IRA can further improve the robustness of speaker extraction model comparing with the models without IRA and more times of IRA can yield more robust model. In $L=8$ case, DPRNN-Spe-IRA leads to 9.78\% improvement of SI-SDR improvement comparing with DPRNN-Spe, and 11.48\% comparing with SpEx+. When $L$ is smaller than $8$, we need to consider an adjustment of ResNets' modeling, which is our future work. Thus, the finer the network, IRA can obtain the greater improvement.

\begin{table}[t!]
\caption{SDRi (dB), SI-SDRi (dB) and PESQ of extracted speech for different systems on the WHAM!. Best scores are highlighted in \textbf{bold}. $L=16$ in DPRNN-Spe and DPRNN-Spe-IRA.}
% \vspace{-2mm}
\label{WHAM!}
\centering 
\small
\scalebox{1.00}{
\begin{tabular}{lcccc}
\hline
\multicolumn{1}{l|}{\textbf{Algorithms}} & \textbf{Params} & \textbf{SI-SDRi} & \textbf{SDRi} & \textbf{PESQ}  \\ \hline
\multicolumn{1}{l|}{Mixture} &-  &0.00 &0.00 &1.66 \\ \hline
% \multicolumn{5}{c}{Blind Speech Separation}   \\ \hline
% \multicolumn{1}{l|}{Conv-TasNet \cite{luo2019conv}} &5.60M  &12.70 &- &- \\ \hline
% \multicolumn{1}{l|}{Deep CASA \cite{liu2019divide}} &13.80M  &13.40  &13.80 & 2.58 \\ \hline
% \multicolumn{1}{l|}{Denoise deep CASA \cite{liu2020deep}} &14.000  &14.400  &14.700 & 2.630 \\ \hline
% \multicolumn{5}{c}{Target Speaker Extraction}   \\ \hline
\multicolumn{1}{l|}{SpEx+ \cite{ge2020spex+}} &13.30M  &13.12  &13.66 &2.46 \\ \hline
\multicolumn{1}{l|}{DPRNN-Spe} &2.91M  & 13.17 &13.78 &2.48  \\ \hline
\multicolumn{1}{l|}{DPRNN-Spe-IRA} &2.94M &\textbf{14.15} &\textbf{14.61} &\textbf{2.57} \\ \hline
\end{tabular}
}
\vspace{-3mm}
\end{table}

We further compare the robustness of DPRNN-Spe and DPRNN-Spe-IRA by the number of bad cases (with SI-SDR lower than 15dB) in closed (validation data) and open (test data) condition under $L=16$, separately in Figure \ref{fig:Distribution}. In closed condition, average SI-SDR is 19.41 dB and 19.51 dB for DPRNN-Spe and DPRNN-Spe-IRA, respectively. Obviously, the number of the bad cases for DPRNN-Spe-IRA is smaller than DPRNN-Spe in almost all low dB ranges. In open condition, average SI-SDR is 17.65 dB and 18.79 dB for DPRNN-Spe and DPRNN-Spe-IRA, respectively. Furthermore, the number of bad cases for DPRNN-Spe becomes bigger than DPRNN-Spe-IRA. In other words, in a matched or unmatched environment, IRA can both improve robustness of DPRNN-Spe by reducing bad cases and IRA can get greater improvement under open condition. As we all know, open condition task is more challenging than closed condition task. Thus, IRA has great application significance for real scenes.

\vspace{-2mm}
\subsection{Comparative study on WHAM!}
\label{wham}

We further verify whether DPRNN-Spe and DPRNN-Spe-IRA are profitable in a noisy environment. Table \ref{WHAM!} illustrates than DPRNN-Spe-IRA achieves 7.44\% relative improvements in terms of SI-SDR over DPRNN-Spe and 7.85\% of SpEx+ (Our implement is based on \cite{Gemengtju2020}).

In summary, the proposed IRA can further improve the robustness of the speaker extraction model, whether in noise-free scenarios or noisy environment.

%\vspace{-2mm}
\section{Conclusions}
\label{sec:conclusions}
This paper introduces a novel iterative refined adaptation strategy for robust speaker extraction tasks. We propose to use a smaller extraction network DPRNN-Spe. Then, iterative refined adaptation is involved to improve the extraction performance. IRA can be applied to any extraction networks easily. Experimental results confirm that IRA makes the extraction network more robust for both noise-free and noisy environments. As future works, we shall explore IRA on more networks and reduce the complexity of models with IRA.

\bibliographystyle{IEEEbib}
\bibliography{refs}

\end{document}